\documentclass[12pt]{iopart}

\usepackage{iopams}
\usepackage{graphicx}

\begin{document}

\title[Observing the nonclassical nature of ultra-broadband bi-photons at ultrafast speed]{Observing the nonclassical nature of ultra-broadband bi-photons at ultrafast speed}

\author{Yaakov Shaked, Roey Pomerantz, Rafi Z. Vered and Avi Pe'er}

\address{Department of physics and BINA Center of nano-technology, Bar-Ilan university, Ramat-Gan 52900, Israel}
\ead{avi.peer@biu.ac.il}
\begin{abstract}
We observe at record-high speed the nonclassical nature of ultra-broadband bi-photons, reducing the measurement time by four orders of magnitude compared to standard techniques of Hong-Ou-Mandel interference or sum-frequency generation. We measure the quantum state of the broadband bi-photons, amplitude and phase, with a pairwise "Mach-Zehnder" quantum interferometer, where bi-photons that are generated in one nonlinear crystal are enhanced (constructive interference) or diminished (destructive interference) in another crystal, depending on the bi-photon phase. We verify the quantum nature of the interference by observing the dependence of the fringe visibility on internal loss. Since destructive interference is equivalent to an attempt to annihilate in the second crystal (by up-conversion) the bi-photons that were created in the first crystal (by down-conversion), the fringe visibility is a measure for the quantum bi-photon purity of the broadband light. The measurement speed-up is due to the large homodyne-like gain from the strong pump ($\!\sim\!10^{7-9}$) in the up-conversion efficiency of single bi-photons, which enables the use of simple photo-detection of the full, ultra-high photon flux instead of single-photon / coincidence counting.
\end{abstract}


\maketitle

\section{Introduction}
Due to quantum correlation, the state of a bi-photon (entangled photon pair) is defined well beyond the uncertainty regarding each of the constituent photons. The inherent quantum nature of bi-photons is a foundation in quantum optics, exploited for many experiments and applications, such as verification of quantum theory \cite{AspectEPR1982,KimbleEPR1992, ZeilingerEPR1995,BoydEPR2004}, engineering of Bell states for quantum information \cite{ZeilingerBellTh1990,BriegelEntaglement2001,EisenbergEntanglement2012,MenicucciQuantumComputing2008,PysherEntanglement2011} and sources of squeezed light for measurements of optical phase below the shot-noise limit \cite{CavesQuantumInterferometer1981,HollandBurnettHeisenbergLimit1993,HollandHeisenbergLimited1998}. A most pronounced realization of this quantum correlation is with ultra-broadband time-energy entangled bi-photons, produced from a narrow pump laser by type-I spontaneous parametric down conversion (SPDC). The precise energy-sum correlation of broadband bi-photons can extend over nearly an octave (more than $100$THz in this report), and their time-difference correlation can be in the few fs regime \cite{HongOuMandelInterferometer1987, AbramCoherence1986, HarrisBroadbandBiPhotons2007, DayanBroadbandBiPhotons2004, AviShapingBiPhotons2005}, thereby providing an extreme realization of the Einstein-Podolsky-Rosen paradox in its original continuous-variable form \cite{EPR1935}. With this ultrashort time correlation, an ultra-high flux of single bi-photons (up to $10^{14}$/s with our configuration) can be generated with negligible probability of multiple pairs \cite{HarrisBroadbandBiPhotons2007, DayanBroadbandBiPhotons2004, AviShapingBiPhotons2005,BarakBroadbandBiphotons2005}.

In spite of their unique quantum properties, broadband bi-photons are the 'black sheep of the family' in current quantum information, and are rarely used in experiments, mainly because of the bandwidth incompetency between the bi-photons and the photo-detectors in standard detection schemes. When the single photons are directly detected, the photo-detectors response time is far too slow ($\!\sim\!100$ps with the fastest available detectors) to resolve the ultrafast correlation time (of order $10-100$fs), and the maximum detectable flux for standard coincidence circuits is limited to few $10^6$ photons/s, inhibiting utilization of the ultra-high flux offered by broadband bi-photons. In frequency domain, this bandwidth incompetency leads to an undesired distinguishability between different frequency pairs of the bi-photons spectrum. Thus, the broad bandwidth of bi-photons is a burden for standard detection, not a resource, and much effort is invested in current experiments to eliminate time-energy entanglement altogether by matching the bi-photons bandwidth to that of the pump, either by narrowing the bi-photons \cite{ChristSingleBIPhotons2012, ChristPureBIPhotons2009, HuangSinglePhotons2011} or by increasing the pump bandwidth using ultrashort pump pulses \cite{GriceBroadPump1997, MosleyPureSingleBiPhotons2008, HodelinBroadbandPump2006}.

In order to fully exploit the bandwidth resource of bi-photons, a different route for detection is required, where the frequency pairs of the bi-photons remain undistinguished. Two major methods were employed so far to address broadband entangled photon pairs - Hong-Ou-Mandel (HOM) interference \cite{HongOuMandelInterferometer1987} and sum-frequency generation (SFG) \cite{BarakBroadbandBiphotons2005, HarrisBroadbandBiPhotons2007, GeorgiadesQuantumSFG1995}. For transform-limited bi-photons both HOM and SFG allow measurement of the ultrashort correlation time, although, the detected photon flux is severely limited in both, either by the use of coincidence detection in HOM, or by the inherently low efficiency of SFG at the single photon level ($\!\sim\! 10^{-10}-10^{-8}$), which yields a very low flux of SFG photons. Both methods are therefore inherently slow and incapable of exploiting the ultrahigh flux.

Since both HOM and SFG are broadband interference effects, both are highly sensitive to spectral phase modulation of the bi-photon input (in somewhat different ways) \cite{AviShapingBiPhotons2005, DayanShapingBiPhotons2007}. Only if the bi-photons are known a-priori to be transform limited can the HOM interferogram provide the bi-photon correlation time, and with SFG, the detection efficiency is strongly hampered by spectral-phase variations, allowing detection only of nearly transform limited bi-photons. By homodyne measurement of the SFG signal against the pump laser, the overall bi-photon phase can be measured \cite{HarrisBroadbandBiPhotons2007,HarrisEntangledPhotons2009}, but not the spectral phase of the constituent frequency-pairs. Thus, exact dispersion compensation is required in both cases, and both methods cannot offer enough information to unravel a general, non uniform bi-photon spectral phase.

\section{Measurement concept}

To measure the bi-photons phase, a pairwise interference against a stable bi-photon reference is required. We utilize for this purpose a well known interference effect \cite{MandelBIphotonInterferometer1991,KorystovBiPhotonInterferometer2001,MandelBiPhotonInterferometerPRL1991,BurlakovBiPhotonInterference1997}, with a most relevant realization in \cite{ZeilingerBiPhotonInterferometer1994}. In the configuration of \cite{ZeilingerBiPhotonInterferometer1994} bi-photons generated by non-collinear, narrowband SPDC, were reflected back along with the pump field for a second pass through the nonlinear crystal, and the photons flux (signal or idler) was measured afterwards, demonstrating high visibility interference, as the relative phase between the pump and the bi-photons was varied. The observed high fringe contrast was a quantum signature of the interference \cite{RealOrVirtualPhotons1995}. Here, we exploit this pairwise interference in order to measure the spectral phase of ultra-broadband bi-photons and to observe their nonclassical nature with near unity efficiency, thereby fully utilizing the ultra-high flux and speeding the measurement by several orders of magnitude. Specifically, we demonstrate the relation between the observed interference contrast and the purity of the bi-photons quantum state (the fraction of bi-photons in the total photon flux).

\begin{figure}[h]
\includegraphics[width=10cm]{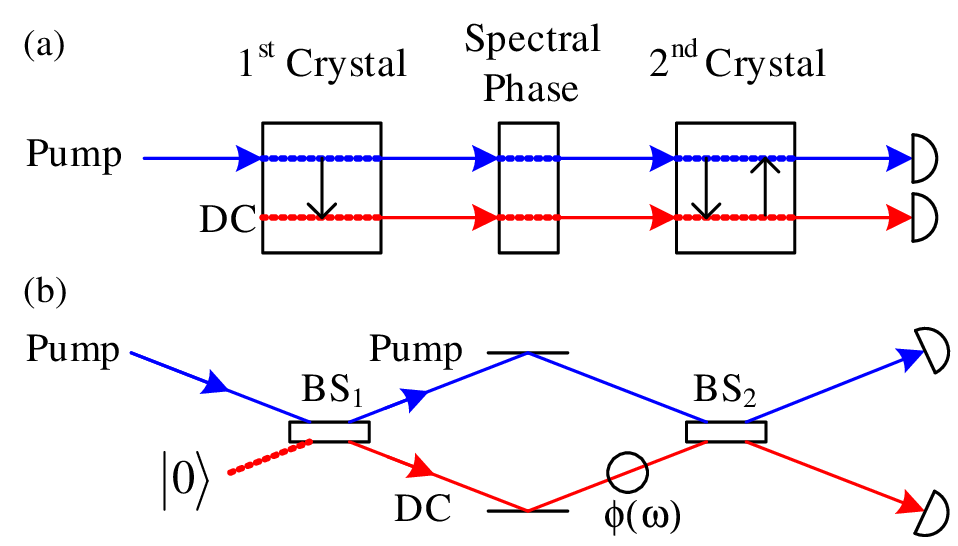}
\caption{\label{Schematic_plus_MachZehnder}(color online) (a) A simplified block diagram of the experiment, showing the generation of bi-photons by SPDC in the first crystal, followed by further enhancement or annihilation of the bi-photons in the second crystal. A change of the relative phase between the pump and the down converted light between the crystals governs the interference. (b) The analogous two-photon Mach-Zehnder, where the crystals represent (unbalanced) beam splitters that couple the pump and the bi-photons beams, allowing a holographic measurement of the bi-photons spectral phase. Loss of photons between the crystals, which reduces the bi-photons state purity is equivalent to an attempt to obtain "which path" information in the interferometer, resulting in a reduced visibility.}
\end{figure}

Our experimental concept is schematically outlined in Fig. \ref{Schematic_plus_MachZehnder}a, where two identical nonlinear crystals in series are pumped by the same pump beam at frequency $\omega_p$ and the bi-photons intensity (or spectrum) is measured after the second crystal. The parametric gain in both crystals is therefore the same, but when bi-photons produced in the first crystal enter the second crystal, they can either enhance further down conversion, or be up-converted back to the pump, depending on the relative phase between the pump and the bi-photons \cite{ZeilingerBiPhotonInterferometer1994,MandelBiPhotonInterferometerPRL1991}. This is a quantum mechanical interference between two indistinguishable possibilities to generate bi-photons - either in the first crystal or in the second. Thus, the described setup is analogous to a Mach-Zehnder interferometer for bi-photons, as illustrated in Fig.\ref{Schematic_plus_MachZehnder}b, where the crystals represent two-photon beam splitters that couple the pump and the down conversion (DC) fields. Conceptually, the 2nd crystal serves as a physical detector of bi-photons, where the existence of an entangled pair is detected by attempting to annihilate it via up-conversion. Since up-conversion affects only bi-photons, the fringe contrast is a direct measure of the bi-photon purity (see analytical derivation later on).

If the bi-photons phase varies spectrally (non transform-limited pairs), high-contrast interference fringes would appear on the measured bi-photons spectrum in a symmetric manner around the degeneracy point at $\omega_p/2$, which provides a direct holographic measurement of the bi-photons spectral phase. Note that the spectral phase of the bi-photons $\phi_s\!+\!\phi_i$ is well defined even though each of the constituent photons cannot be assigned a definite phase $\phi_{s,i}$. Thus, the interference can reveal only symmetric phase variations and is insensitive to anti-symmetric phase that keeps the phase-sum intact. Ideally, the bi-photons are born at the first crystal with a flat spectral phase ($\phi_s\!+\!\phi_i\!=\!0$ for all frequencies), but imperfect phase matching in the crystal, dispersion of optical elements in the beam, or deliberate pulse shaping can modify it in many ways.

\section{Experiment and results}
\begin{figure}[h]
\includegraphics[width=10cm]{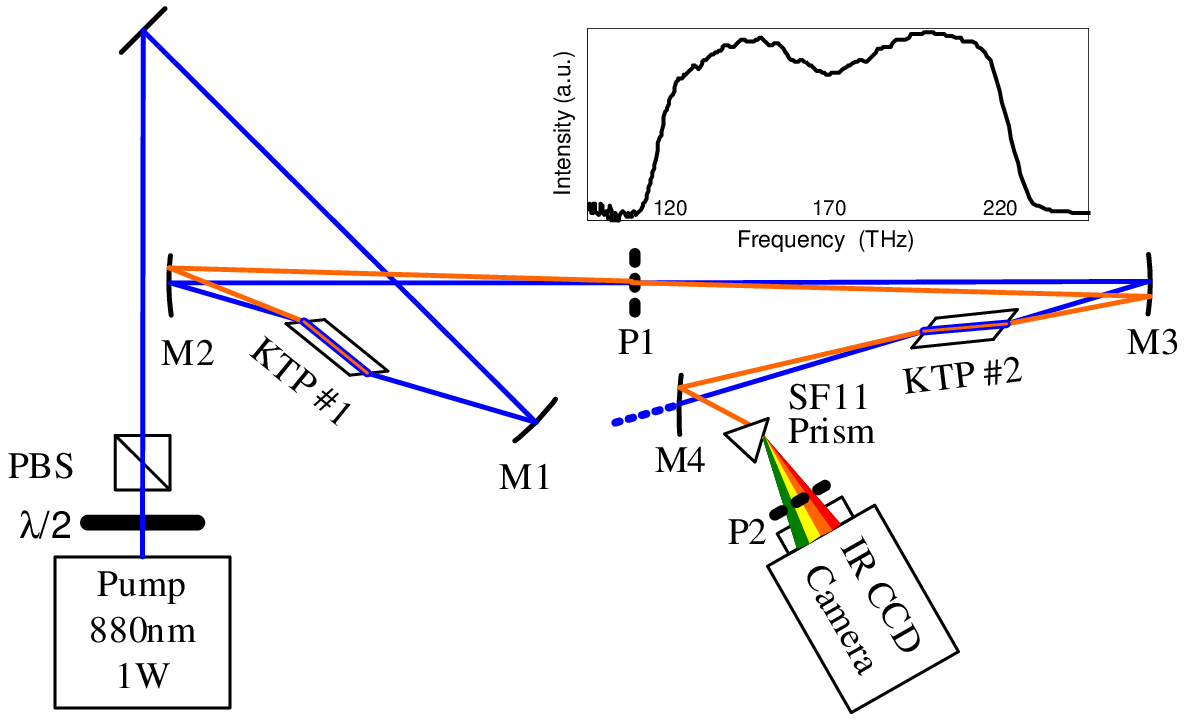}
\caption{\label{setup} (color online) Experimental layout: bi-photons are generated in the 1st crystal ($12 mm$ long PPKTP), pumped by a single-frequency diode laser at $880\ nm$ with up to $1\ W$ power. Reflection of the Bi Photons from mirrors M2 and M3 is accompanied by a spectral phase shift. Both the bi-photons and the pump are directed into a second identical crystal, where further generation of bi-photons or up-conversion back to the pump can occur. The resulting bi-photons spectrum is measured by a home-built spectrometer composed of a prism (SF11) and a CCD camera with 7ms integration time (Xeva-2.5-320 by Xenics). The last mirror (M4) separates the pump from the bi-photons, allowing the pump power to be measured. Attenuation is achieved either before the 1st crystal by a half-wave plate and polarizer, or between the two crystals with a polarizer P1, and a polarizer P2 in front of the camera. The inset shows a measured intensity spectrum of the ultra-broadband bi-photons after the first crystal}
\end{figure}

A simplified layout of our experiment is shown in figure \ref{setup}. In order to demonstrate that the observed fringe contrast is a nonclassical feature that reflects the quantum state purity of the bi-photons, we compare two experimental scenarios, where a classical analysis predicts a difference that is \emph{opposite and drastically different} from the experimental quantum mechanical result: we attenuate the light entering the second crystal in two ways - once by attenuating the pump before the first crystal, thereby reducing also the generated bi-photons flux; and second, by attenuating both the pump and the bi-photons beam between the two crystals. Quantum-mechanically, the first scenario attenuates the spontaneous generation rate of bi-photons (two-photon attenuation), but does not alter their purity, whereas the second scenario attenuates every photon independently (one-photon attenuation), reducing the single photon flux linearly, but the bi-photon flux \emph{quadratically}, thereby diminishing the bi-photon state purity. The bi-photon interference contrast is therefore expected to remain high for two-photon attenuation (first scenario), but will diminish for one-photon attenuation (second scenario). In the Mach-Zehnder analog, attenuation between the crystals is equivalent to an attempt to obtain "which path" information by "stealing" one of the photons, causing the interference contrast to diminish \cite{EnglertWhichWayInformation1996}.

In a classical analysis on the other hand, down conversion is a parametric, phase-dependent amplification process, where spontaneous bi-photon emission is accounted for by assuming an input white noise field as a classical representation for the vacuum. This input noise is later quadrature-squeezed by the parametric gain, which is proportional to the pump amplitude. Since the total parametric gain in the first scenario is reduced quadratically (the pump is attenuated in both crystals) but only linearly in the the second scenario (the pump is attenuated only in one crystal), the classical expectation is that the fringe contrast would decay quadratically for two-photon attenuation (first scenario), but linearly for one-photon attenuation (second scenario) - an \emph{opposite} prediction to the quantum thinking. A detail of both the classical analysis and the complete quantum model is given in section \ref{Theory} below.

\begin{figure}[h]
\includegraphics[width=10cm]{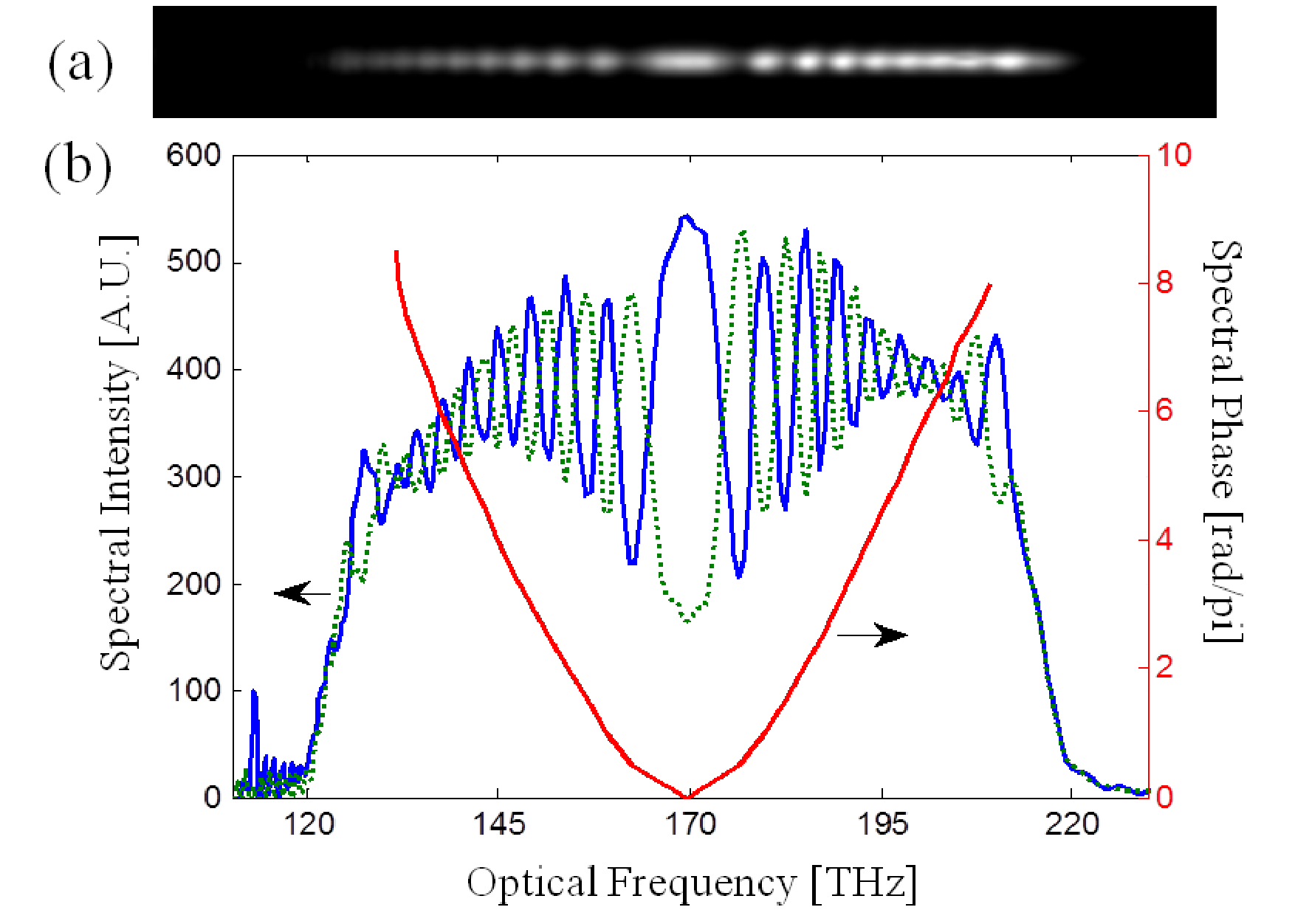}
\caption{\label{spectrum} (color online) Spectral fringes (a) normalized CCD image of the spectral fringes after both crystals. (b) Two calibrated fringe spectra with a  $\pi$ phase shift between them (blue and dotted green lines); and the corresponding calculated spectral phase $\Phi(\omega)/2$ of the bi-photons (red line). }
\end{figure}

In the experiment (Fig. \ref{setup}), a single frequency diode laser at $880nm$ pumps two identical KTP crystals, periodically polled for collinear down conversion around $1760nm$. This pump was chosen to coincide the center of the bi-photons spectrum with the zero-dispersion wavelength of the KTP crystal, allowing an ultra-broad phase matching (over $100THz$, nearly an octave) for collinear down conversion between $1.3-2.5 \mu m$, as illustrated in the inset of Fig.\ref{setup}. Such bandwidth corresponds to a maximum possible bi-photon flux of $F_{max}=1.08\times10^{14}$ photons/s, nearly $12\ \mu W$ of single bi-photons. The actual flux in the experiment was approximately $F_{max}/90$, limited by the available pump power, well within the single bi-photon regime. The down converted light from the first KTP crystal continues along with the pump into the second identical KTP crystal, where either down conversion or up conversion back to the pump can occur, and the down conversion spectrum is measured after the second crystal with a home-built prism-based spectrometer. The spectral modulation of the bi-photons phase is partially due to residual phase mismatch in the crystals and mainly due to negative dispersion of the broadband dielectric mirrors (M2, M3) between the crystals, causing interference fringes to appear on the bi-photons spectrum, as shown in Fig. \ref{spectrum}. We use this fringe pattern to reconstruct the bi-photon phase, as shown in Fig.\ref{spectrum} (red line). A great convenience of our interferometer configuration is its fully collinear arrangement, which renders it insensitive to path length fluctuations. The observed fringe pattern is therefore inherently stable with no active phase locking.
\begin{figure}[h]
\includegraphics[width=10cm]{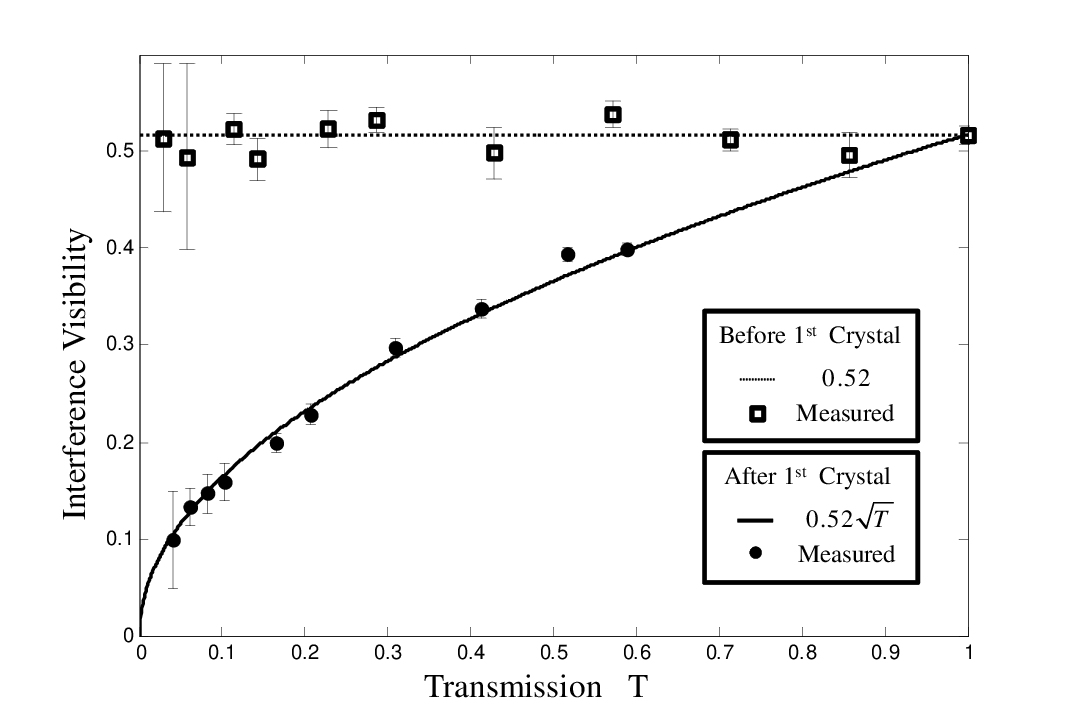}
\caption{\label{contrast} Measured interference visibility as function of attenuation with corresponding theoretical fits before the first crystal (squares + dotted fit) and between crystals (circles + solid fit).}
\end{figure}

By a slight lateral shift of the Brewster-cut PPKTP crystal, the relative phase between the pump and the bi-photons can be scanned. Doing so, we measure the interference visibility across the spectrum, obtaining a maximal visibility of $52\%$ for near degenerate bi-photons, and decreasing to $\!\sim\!20\%$ at the spectrum edges. According to the model laid out hereon, this corresponds to a purity of $\!\sim\! 27\%$. We assume that the two-photon purity in our experiment was actually higher, since the measured contrast was technically limited by non-perfect spatial mode matching across the ultra-broad spectrum between the pump and the down-converted beam.

The effect of the purity of the quantum state on the interference visibility is presented in Fig.\ref{contrast}, by comparing the two loss scenarios mentioned above. Attenuation of the pump before the first crystal has no effect on the interference visibility (Fig. \ref{contrast} squares), whereas attenuating both the pump and SPDC fields between the crystals, results in reduction of the interference visibility (Fig.\ref{contrast} circles) with excellent agreement to the square root dependence of Eq. \ref{sqrtt} below (solid line). The interference visibility provides therefore a method of measuring and monitoring the bi-photon quantum state purity - we confirm the presence of a bi-photon by attempting to annihilate it in the second crystal.

Note that a dark fringe of the spectral interferogram represents up-conversion of single bi-photons back to the pump with $\!>\!50\%$ efficiency (!), orders of magnitude higher than with direct SFG \cite{AviShapingBiPhotons2005, BarakBroadbandBiphotons2005}. This phenomenal enhancement is due to the homodyne-like gain in the two-photon efficiency from the strong pump in the second crystal. Indeed, the generated SFG photons cannot be directly detected on top of the intense pump, but the absence of the down-converted photons is easily measurable by detecting the average photon flux (intensity) with a simple photo-detector. It is critical to note here that we only measure the annihilation \textit{probability} of photons, not the specific annihilation of a specific bi-photon (there are so many of them…). With an average photon flux of $10^{12}$ photons/second (as in our experiment), it is very easy to detect the (average) absence of photons with very close to $100\%$ certainty. Furthermore, the ultra-high flux of bi-photons (due to the ultra-broad bandwidth) allows to measure at high speed not only the average photon flux, but also its spectrum, thereby providing access to the spectral phase of the bi-photons which is inaccessible for HOM or SFG. Specifically, the spectrum was captured over $\sim\!150$ pixels of the CCD camera with very low noise at an integration time of $\sim\!7ms$, indicating an incoming flux of $10^7\!-\!10^8$ photons / pixel / detection time. This should be compared to detection times of $100\!-\!1000s$, required to collect the same number of photons with SFG or HOM coincidence, representing a $\!10^4\!-\!10^5$ speedup. Even faster detection could be achieved for transform-limited bi-photons, where the interference is uniform over the entire spectrum, allowing detection of the full photon flux on a single fast photo-detector.

\section{\label{Theory}Model}
\subsection{Classical Analysis}
Before we dwell into the complete quantum model, it is beneficial to explore the failure of the classical model to account for the results, and the reasons that lead to this failure. Classically, each crystal is represented by a parametric, phase-dependent amplifier, which links the output field to the input field by $A_{out}=\cosh\left(g_j\right)A_{in}+\sinh\left(g_j\right)A^{*}_{in}$, where $g_j=\chi|A_{p,j}|$ is the parametric gain in crystal $j$ proportional to the pump amplitude. Let us first consider the process in the second crystal: Assuming an input of two quadratures $A_{in,2}=\alpha+i\beta$, the output field becomes $A_{out,2}=\alpha e^{g_2}+i\beta e^{-g_2}$, and the output intensity is $I=\alpha^2 e^{2g_2}+\beta^2 e^{-2g_2}$, indicating that one input quadrature is amplified and the other attenuated by $g_2$. If the pump phase is switched, the roles of the quadratures is also switched, and the output intensity becomes $I'=\alpha^2 e^{-2g_2}+\beta^2 e^{2g_2}$. The expected classical interference contrast is therefore
\begin{equation}
V_{classical}=\frac{I-I'}{I+I'}=\frac{\left(\alpha^2-\beta^2\right)\left(e^{2g_2}-e^{-2g_2}\right)}{\left(\alpha^2+\beta^2\right)\left(e^{2g_2}+e^{-2g_2}\right)}
                =\frac{\alpha^2-\beta^2}{\alpha^2+\beta^2}\tanh(2g_2).
\end{equation}
Evidently, the classical contrast depends on both the parametric gain $g_2$ and on the "classical squeezing ratio" $\left(\alpha^2-\beta^2\right)/\left(\alpha^2+\beta^2\right)$ at the input to the second crystal. In a classical regime of high gain, the contrast at the output of the second crystal may serve as a measure for the level of quadratures squeezing at its input.

The input to the second crystal is the output of the first crystal, which can be written as $A_{in,2}=A_{out,1}=\alpha_0 e^{g_1}+i\beta_0 e^{-g_1}$, where $\alpha_0, \beta_0$ are real, white-noise random variables that represent the two input quadratures with equal average intensity $\left\langle\alpha^{2}_0\right\rangle=\left\langle\beta^{2}_0\right\rangle=n$ (we assume the input to the first crystal to be unsqueezed noise). The "classical squeezing ratio" at the output of the first crystal is therefore
\begin{equation}
\frac{\left\langle\alpha^{2}_0\right\rangle e^{2g_1}-\left\langle\beta^{2}_0\right\rangle e^{-2g_1}}{\left\langle\alpha^{2}_0\right\rangle e^{2g_1}+\left\langle\beta^{2}_0\right\rangle e^{-2g_1}}=\tanh(2g_1),
\end{equation}
indicating that the overall classical contrast should be
\begin{equation}
V_{classical}=\tanh(2g_1)\tanh(2g_2)\approx4 g_1 g_2\propto\left|A_{p,1}A_{p,2}\right|=\sqrt{I_{p,1}I_{p,2}},
\end{equation}
where a low gain approximation $\tanh(g_j)\approx g_j\propto \sqrt{I_{p,j}}$ was assumed for both crystals.

The classical prediction for the visibility is now clear for both scenarios. For two-photon attenuation (first scenario), the gain in both crystals is equal ($g_1=g_2=g$), and the visibility $V_{classical}\propto g^2=\left|A_{p}\right|^2$ should decay linearly with the pump intensity, and faster than the one-photon attenuation scenario. This is in contradiction with the experiment (and quantum model below), that show an intensity independent contrast for two-photon attenuation. The classical prediction fails here mainly because of the inability to model classically the vacuum input of the first crystal at low pump intensities. It is surprising, however that for one-photon attenuation (second scenario), this naive classical model does yield the correct square-root scaling for the contrast, even at very low intensities (since only the gain in the second crystal is attenuated, a square root dependence on the pump intensity is expected).

\subsection{Quantum Model}
To fully account for the observed results, a quantum model is necessary. Due to the low efficiency of SPDC we can neglect multiple-pairs, and assume a perturbative propagator of the form $U\!=\!e^{iHt}\!\approx\!1+iHt\!=\!1+\alpha_{\omega} a_{\omega}^{\dagger} a_{-\omega}^{\dagger}$, where the creation operators generate a photon in the modes $\omega_p/2+\omega$ and $\omega_p/2-\omega$. The coefficient $\alpha_\omega$ is a weighted probability amplitude for generating a photon-pair $\left| 1_{\omega} , 1_{-\omega} \right\rangle$, assumed small. Assuming a vacuum $\left| 0 \right\rangle$ input before the first crystal, the quantum state after it is
\begin{equation}
\left| \psi  \right\rangle_{_{1}}  = \left| 0 \right\rangle + \alpha_{\omega} \left| 1_{\omega} , 1_{-\omega} \right\rangle.
\end{equation}
Loss between the two crystals can be modeled as a beam splitter (BS), with reflection (absorption) and transmission amplitude coefficients $r$ and $t$,  positioned between the crystals, which mixes the bi-photons from the first crystal with an additional vacuum state $\left| 0 \right\rangle_2$ from its other port. Propagating the output state from the first crystal through the beam splitter yields
\begin{equation}
\label{BSoutput}
\left| \psi  \right\rangle _{_{BS}}\!=\!\left| 0 \right\rangle _1 \left| 0 \right\rangle _2\!+\!\alpha_{\omega}\!\left[\!\begin{array}{l}
 t^2 \left| {1_\omega  ,1_{ - \omega } } \right\rangle _1 \left| 0 \right\rangle _2    \\
 -r^2 \left| 0 \right\rangle _1 \left| {1_\omega  ,1_{ - \omega } } \right\rangle _2    \\
 +irt\left| {1_\omega  ,0_{ - \omega } } \right\rangle _1 \left| {0_\omega  ,1_{ - \omega } } \right\rangle _2    \\
 +irt\left| {0_\omega  ,1_{ - \omega } } \right\rangle _1 \left| {1_\omega  ,0_{ - \omega } } \right\rangle _2  \\
 \end{array}\!\right].
\end{equation}
The ket indices 1,2 stand for the transmitted port and the loss port respectively. The four terms in eq. \ref{BSoutput} represent the four possibilities for the pair after the loss BS (fully transmitted, fully reflected and two possibilities of one transmitted + one reflected).

The second crystal is positioned after a frequency dependent phase was acquired and the pump power entering the crystal is also attenuated by the loss. The second crystal propagator is therefore
\begin{equation}
\left( {1 + t\alpha _\omega  e^{i\Phi \left( \omega  \right)} a_\omega ^{ + \left( 1 \right)} a_{ - \omega }^{ + \left( 1 \right)} } \right),
\end{equation}
where $\Phi \left( \omega  \right)=\phi \left( \omega_p/2 + \omega \right) + \phi \left(\omega_p/2 - \omega \right)$ is the relative phase between the frequency pair and the pump. The propagated state after the 2nd crystal is
\begin{equation}
\left|\psi\right\rangle_{_{2}}\!=\!\left|0\right\rangle _1\left|0\right\rangle_2\!+\!\alpha_{\omega} \left[ \begin{array}{l}
 \left( {t^2  + te^{i\Phi \left( \omega  \right)} } \right)\left| {1_\omega  ,1_{ - \omega } } \right\rangle _1 \left| 0 \right\rangle _2    \\
 -r^2 \left| 0 \right\rangle _1 \left| {1_\omega  ,1_{ - \omega } } \right\rangle _2    \\
 +irt\left| {1_\omega  ,0_{ - \omega } } \right\rangle _1 \left| {0_\omega  ,1_{ - \omega } } \right\rangle _2    \\
 +irt\left| {0_\omega  ,1_{ - \omega } } \right\rangle _1 \left| {1_\omega  ,0_{ - \omega } } \right\rangle _2  \\
 \end{array}\!\right],
\end{equation}
and the detected intensity at the transmitted port is
\begin{equation}
\!I_\omega\!\propto\!\left\langle {\psi _f } \right|a_\omega ^{ + \left( 1 \right)} a_\omega ^{\left( 1 \right)} \left| {\psi _f } \right\rangle\!=\!\left| {\alpha _\omega  } \right|^2\!\left| t \right|^2\!\left\{ {1\!+\!t\cos \left( {\Phi \left( \omega  \right)} \right)} \right\}.
\end{equation}
The visibility $ V\!=\!{{\left( {I_{\omega ,\max }\!-\!I_{\omega ,\min }}\right)} \mathord{\left/
 {\vphantom {{\left( {I_{\omega ,\max }\!-\!I_{\omega ,\min}}\right)}{\left({I_{\omega ,\max }\!+\!I_{\omega ,\min}}\right)}}} \right.
 \kern-\nulldelimiterspace} {\left( {I_{\omega ,\max }\!+\!I_{\omega ,\min } } \right)}} $ depends therefore on the loss as
\begin{equation} \label{sqrtt}
V(t) = \left| t \right| = \sqrt{T},
\end{equation}
indicating a linear proportion to the amplitude transmission, as indeed observed in Fig. \ref{contrast}. The purity of the quantum bi-photon state - the fraction of entangled photons at a specific frequency out of the total number of photons at that specific frequency ($\eta\left(t \right)\!\equiv\!\left\langle N^{pairs}\right\rangle\!/\!\left\langle N^{photons} \right\rangle$), is directly related to the loss by $\eta \left( t  \right)\!=\!\left| t \right|^2\!=\!V^2$. The visibility of the fringes therefore directly reflects the purity.

\section{Conclusion}
We carried out an 'ultrafast' measurement of the complete quantum wave-function of ultra-broadband bi-photons by using a pairwise quantum interferometer. We demonstrated the square root dependence of the interference contrast on loss transmission, and derived the relation between the purity of the bi-photon state and the observed fringe contrast. The high two-photon efficiency at the single bi-photons level, enhanced orders of magnitude by the intense pump (ideally to unity) and the ultra-high flux of bi-photons, speed and simplify the measurement considerably, allowing observation of the bi-photons non-classicality \emph{at a detection rate comparable to the photon flux}. We expect this method to become an important member of the quantum optics toolbox for broadband time-energy entangled photons.

\ack{This research was supported by the EU-IRG program (grant no. 248630).}

\Bibliography{34}

\expandafter\ifx\csname natexlab\endcsname\relax\def\natexlab#1{#1}\fi
\expandafter\ifx\csname bibnamefont\endcsname\relax
  \def\bibnamefont#1{#1}\fi
\expandafter\ifx\csname bibfnamefont\endcsname\relax
  \def\bibfnamefont#1{#1}\fi
\expandafter\ifx\csname citenamefont\endcsname\relax
  \def\citenamefont#1{#1}\fi
\expandafter\ifx\csname url\endcsname\relax
  \def\url#1{\texttt{#1}}\fi
\expandafter\ifx\csname urlprefix\endcsname\relax\def\urlprefix{URL }\fi
\providecommand{\bibinfo}[2]{#2}
\providecommand{\eprint}[2][]{\url{#2}}

\bibitem{AspectEPR1982}
\bibinfo{author}{\bibfnamefont{A.}~\bibnamefont{Aspect}},
  \bibinfo{author}{\bibfnamefont{P.}~\bibnamefont{Grangier}}, \bibnamefont{and}
  \bibinfo{author}{\bibfnamefont{G.}~\bibnamefont{Roger}},
  \bibinfo{journal}{Phys. Rev. Lett.} \textbf{\bibinfo{volume}{49}},
  \bibinfo{pages}{91} (\bibinfo{year}{1982}).

\bibitem{KimbleEPR1992}
\bibinfo{author}{\bibfnamefont{Z.}~\bibnamefont{Ou}},
  \bibinfo{author}{\bibfnamefont{S.}~\bibnamefont{Pereira}}, \bibnamefont{and}
  \bibinfo{author}{\bibfnamefont{H.}~\bibnamefont{Kimble}},
  \bibinfo{journal}{Applied Physics B} \textbf{\bibinfo{volume}{55}},
  \bibinfo{pages}{265} (\bibinfo{year}{1992}), ISSN \bibinfo{issn}{0946-2171}.

\bibitem{ZeilingerEPR1995}
\bibinfo{author}{\bibfnamefont{P.~G.} \bibnamefont{Kwiat}},
  \bibinfo{author}{\bibfnamefont{K.}~\bibnamefont{Mattle}},
  \bibinfo{author}{\bibfnamefont{H.}~\bibnamefont{Weinfurter}},
  \bibinfo{author}{\bibfnamefont{A.}~\bibnamefont{Zeilinger}},
  \bibinfo{author}{\bibfnamefont{A.~V.} \bibnamefont{Sergienko}},
  \bibnamefont{and} \bibinfo{author}{\bibfnamefont{Y.}~\bibnamefont{Shih}},
  \bibinfo{journal}{Phys. Rev. Lett.} \textbf{\bibinfo{volume}{75}},
  \bibinfo{pages}{4337} (\bibinfo{year}{1995}).

\bibitem{BoydEPR2004}
\bibinfo{author}{\bibfnamefont{J.~C.} \bibnamefont{Howell}},
  \bibinfo{author}{\bibfnamefont{R.~S.} \bibnamefont{Bennink}},
  \bibinfo{author}{\bibfnamefont{S.~J.} \bibnamefont{Bentley}},
  \bibnamefont{and} \bibinfo{author}{\bibfnamefont{R.~W.} \bibnamefont{Boyd}},
  \bibinfo{journal}{Phys. Rev. Lett.} \textbf{\bibinfo{volume}{92}},
  \bibinfo{pages}{210403} (\bibinfo{year}{2004}).

\bibitem{ZeilingerBellTh1990}
\bibinfo{author}{\bibfnamefont{D.~M.} \bibnamefont{Greenberger}},
  \bibinfo{author}{\bibfnamefont{M.~A.} \bibnamefont{Horne}},
  \bibinfo{author}{\bibfnamefont{A.}~\bibnamefont{Shimony}}, \bibnamefont{and}
  \bibinfo{author}{\bibfnamefont{A.}~\bibnamefont{Zeilinger}},
  \bibinfo{journal}{American Journal of Physics} \textbf{\bibinfo{volume}{58}},
  \bibinfo{pages}{1131} (\bibinfo{year}{1990}).

\bibitem{BriegelEntaglement2001}
\bibinfo{author}{\bibfnamefont{H.~J.} \bibnamefont{Briegel}} \bibnamefont{and}
  \bibinfo{author}{\bibfnamefont{R.}~\bibnamefont{Raussendorf}},
  \bibinfo{journal}{Phys. Rev. Lett.} \textbf{\bibinfo{volume}{86}},
  \bibinfo{pages}{910} (\bibinfo{year}{2001}).

\bibitem{EisenbergEntanglement2012}
\bibinfo{author}{\bibfnamefont{E.}~\bibnamefont{Megidish}},
  \bibinfo{author}{\bibfnamefont{T.}~\bibnamefont{Shacham}},
  \bibinfo{author}{\bibfnamefont{A.}~\bibnamefont{Halevy}},
  \bibinfo{author}{\bibfnamefont{L.}~\bibnamefont{Dovrat}}, \bibnamefont{and}
  \bibinfo{author}{\bibfnamefont{H.~S.} \bibnamefont{Eisenberg}},
  \bibinfo{journal}{Phys. Rev. Lett.} \textbf{\bibinfo{volume}{109}},
  \bibinfo{pages}{080504} (\bibinfo{year}{2012}).

\bibitem{MenicucciQuantumComputing2008}
\bibinfo{author}{\bibfnamefont{N.~C.} \bibnamefont{Menicucci}},
  \bibinfo{author}{\bibfnamefont{S.~T.} \bibnamefont{Flammia}},
  \bibnamefont{and} \bibinfo{author}{\bibfnamefont{O.}~\bibnamefont{Pfister}},
  \bibinfo{journal}{Phys. Rev. Lett.} \textbf{\bibinfo{volume}{101}},
  \bibinfo{pages}{130501} (\bibinfo{year}{2008}).

\bibitem{PysherEntanglement2011}
\bibinfo{author}{\bibfnamefont{M.}~\bibnamefont{Pysher}},
  \bibinfo{author}{\bibfnamefont{Y.}~\bibnamefont{Miwa}},
  \bibinfo{author}{\bibfnamefont{R.}~\bibnamefont{Shahrokhshahi}},
  \bibinfo{author}{\bibfnamefont{R.}~\bibnamefont{Bloomer}}, \bibnamefont{and}
  \bibinfo{author}{\bibfnamefont{O.}~\bibnamefont{Pfister}},
  \bibinfo{journal}{Phys. Rev. Lett.} \textbf{\bibinfo{volume}{107}},
  \bibinfo{pages}{030505} (\bibinfo{year}{2011}).

\bibitem{CavesQuantumInterferometer1981}
\bibinfo{author}{\bibfnamefont{C.~M.} \bibnamefont{Caves}},
  \bibinfo{journal}{Phys. Rev. D} \textbf{\bibinfo{volume}{23}},
  \bibinfo{pages}{1693} (\bibinfo{year}{1981}).

\bibitem{HollandBurnettHeisenbergLimit1993}
\bibinfo{author}{\bibfnamefont{M.~J.} \bibnamefont{Holland}} \bibnamefont{and}
  \bibinfo{author}{\bibfnamefont{K.}~\bibnamefont{Burnett}},
  \bibinfo{journal}{Phys. Rev. Lett.} \textbf{\bibinfo{volume}{71}},
  \bibinfo{pages}{1355} (\bibinfo{year}{1993}).

\bibitem{HollandHeisenbergLimited1998}
\bibinfo{author}{\bibfnamefont{T.}~\bibnamefont{Kim}},
  \bibinfo{author}{\bibfnamefont{O.}~\bibnamefont{Pfister}},
  \bibinfo{author}{\bibfnamefont{M.~J.} \bibnamefont{Holland}},
  \bibinfo{author}{\bibfnamefont{J.}~\bibnamefont{Noh}}, \bibnamefont{and}
  \bibinfo{author}{\bibfnamefont{J.~L.} \bibnamefont{Hall}},
  \bibinfo{journal}{Phys. Rev. A} \textbf{\bibinfo{volume}{57}},
  \bibinfo{pages}{4004} (\bibinfo{year}{1998}).

\bibitem{HongOuMandelInterferometer1987}
\bibinfo{author}{\bibfnamefont{C.~K.} \bibnamefont{Hong}},
  \bibinfo{author}{\bibfnamefont{Z.~Y.} \bibnamefont{Ou}}, \bibnamefont{and}
  \bibinfo{author}{\bibfnamefont{L.}~\bibnamefont{Mandel}},
  \bibinfo{journal}{Phys. Rev. Lett.} \textbf{\bibinfo{volume}{59}},
  \bibinfo{pages}{2044} (\bibinfo{year}{1987}).

\bibitem{AbramCoherence1986}
\bibinfo{author}{\bibfnamefont{I.}~\bibnamefont{Abram}},
  \bibinfo{author}{\bibfnamefont{R.~K.} \bibnamefont{Raj}},
  \bibinfo{author}{\bibfnamefont{J.~L.} \bibnamefont{Oudar}}, \bibnamefont{and}
  \bibinfo{author}{\bibfnamefont{G.}~\bibnamefont{Dolique}},
  \bibinfo{journal}{Phys. Rev. Lett.} \textbf{\bibinfo{volume}{57}},
  \bibinfo{pages}{2516} (\bibinfo{year}{1986}).

\bibitem{HarrisBroadbandBiPhotons2007}
\bibinfo{author}{\bibfnamefont{S.~E.} \bibnamefont{Harris}},
  \bibinfo{journal}{Phys. Rev. Lett.} \textbf{\bibinfo{volume}{98}},
  \bibinfo{pages}{063602} (\bibinfo{year}{2007}).

\bibitem{DayanBroadbandBiPhotons2004}
\bibinfo{author}{\bibfnamefont{B.}~\bibnamefont{Dayan}},
  \bibinfo{author}{\bibfnamefont{A.}~\bibnamefont{Pe'er}},
  \bibinfo{author}{\bibfnamefont{A.~A.} \bibnamefont{Friesem}},
  \bibnamefont{and}
  \bibinfo{author}{\bibfnamefont{Y.}~\bibnamefont{Silberberg}},
  \bibinfo{journal}{Phys. Rev. Lett.} \textbf{\bibinfo{volume}{93}},
  \bibinfo{pages}{023005} (\bibinfo{year}{2004}).

\bibitem{AviShapingBiPhotons2005}
\bibinfo{author}{\bibfnamefont{A.}~\bibnamefont{Pe'er}},
  \bibinfo{author}{\bibfnamefont{B.}~\bibnamefont{Dayan}},
  \bibinfo{author}{\bibfnamefont{A.~A.} \bibnamefont{Friesem}},
  \bibnamefont{and}
  \bibinfo{author}{\bibfnamefont{Y.}~\bibnamefont{Silberberg}},
  \bibinfo{journal}{Phys. Rev. Lett.} \textbf{\bibinfo{volume}{94}},
  \bibinfo{pages}{073601} (\bibinfo{year}{2005}).

\bibitem{BarakBroadbandBiphotons2005}
\bibinfo{author}{\bibfnamefont{B.}~\bibnamefont{Dayan}},
  \bibinfo{author}{\bibfnamefont{A.}~\bibnamefont{Pe'er}},
  \bibinfo{author}{\bibfnamefont{A.~A.} \bibnamefont{Friesem}},
  \bibnamefont{and}
  \bibinfo{author}{\bibfnamefont{Y.}~\bibnamefont{Silberberg}},
  \bibinfo{journal}{Phys. Rev. Lett.} \textbf{\bibinfo{volume}{94}},
  \bibinfo{pages}{043602} (\bibinfo{year}{2005}).

\bibitem{EPR1935}
\bibinfo{author}{\bibfnamefont{A.}~\bibnamefont{Einstein}},
  \bibinfo{author}{\bibfnamefont{B.}~\bibnamefont{Podolsky}},
    \bibnamefont{and}
  \bibinfo{author}{\bibfnamefont{N.}~\bibnamefont{Rosen}},
  \bibinfo{journal}{Phys. Rev.} \textbf{\bibinfo{volume}{41}},
  \bibinfo{pages}{777} (\bibinfo{year}{1935}).

\bibitem{DayanShapingBiPhotons2007}
\bibinfo{author}{\bibfnamefont{B.}~\bibnamefont{Dayan}},
  \bibinfo{author}{\bibfnamefont{Y.}~\bibnamefont{Bromberg}},
  \bibinfo{author}{\bibfnamefont{I.}~\bibnamefont{Afek}},
  \bibnamefont{and}
  \bibinfo{author}{\bibfnamefont{Y.}~\bibnamefont{Silberberg}},
  \bibinfo{journal}{Phys. Rev. A} \textbf{\bibinfo{volume}{75}},
  \bibinfo{pages}{043804} (\bibinfo{year}{2007}).

  \bibitem{ChristSingleBIPhotons2012}
\bibinfo{author}{\bibfnamefont{A.}~\bibnamefont{Christ}} \bibnamefont{and}
  \bibinfo{author}{\bibfnamefont{C.}~\bibnamefont{Silberhorn}},
  \bibinfo{journal}{Phys. Rev. A} \textbf{\bibinfo{volume}{85}},
  \bibinfo{pages}{023829} (\bibinfo{year}{2012}).

\bibitem{ChristPureBIPhotons2009}
\bibinfo{author}{\bibfnamefont{A.}~\bibnamefont{Christ}},
  \bibinfo{author}{\bibfnamefont{A.}~\bibnamefont{Eckstein}},
  \bibinfo{author}{\bibfnamefont{P.~J.} \bibnamefont{Mosley}},
  \bibnamefont{and}
  \bibinfo{author}{\bibfnamefont{C.}~\bibnamefont{Silberhorn}},
  \bibinfo{journal}{Opt. Express} \textbf{\bibinfo{volume}{17}},
  \bibinfo{pages}{3441} (\bibinfo{year}{2009}).

\bibitem{HuangSinglePhotons2011}
\bibinfo{author}{\bibfnamefont{Y.-P.} \bibnamefont{Huang}},
  \bibinfo{author}{\bibfnamefont{J.~B.} \bibnamefont{Altepeter}},
  \bibnamefont{and} \bibinfo{author}{\bibfnamefont{P.}~\bibnamefont{Kumar}},
  \bibinfo{journal}{Phys. Rev. A} \textbf{\bibinfo{volume}{84}},
  \bibinfo{pages}{033844} (\bibinfo{year}{2011}).

\bibitem{GriceBroadPump1997}
\bibinfo{author}{\bibfnamefont{W.~P.} \bibnamefont{Grice}} \bibnamefont{and}
  \bibinfo{author}{\bibfnamefont{I.~A.} \bibnamefont{Walmsley}},
  \bibinfo{journal}{Phys. Rev. A} \textbf{\bibinfo{volume}{56}},
  \bibinfo{pages}{1627} (\bibinfo{year}{1997}).

\bibitem{MosleyPureSingleBiPhotons2008}
\bibinfo{author}{\bibfnamefont{P.~J.} \bibnamefont{Mosley}},
  \bibinfo{author}{\bibfnamefont{J.~S.} \bibnamefont{Lundeen}},
  \bibinfo{author}{\bibfnamefont{B.~J.} \bibnamefont{Smith}},
  \bibinfo{author}{\bibfnamefont{P.}~\bibnamefont{Wasylczyk}},
  \bibinfo{author}{\bibfnamefont{A. B.} \bibnamefont{URen}},
  \bibinfo{author}{\bibfnamefont{C.}~\bibnamefont{Silberhorn}},
  \bibnamefont{and} \bibinfo{author}{\bibfnamefont{I.~A.}
  \bibnamefont{Walmsley}}, \bibinfo{journal}{Phys. Rev. Lett.}
  \textbf{\bibinfo{volume}{100}}, \bibinfo{pages}{133601}
  (\bibinfo{year}{2008}).

\bibitem{HodelinBroadbandPump2006}
\bibinfo{author}{\bibfnamefont{J.~F.} \bibnamefont{Hodelin}},
  \bibinfo{author}{\bibfnamefont{G.}~\bibnamefont{Khoury}}, \bibnamefont{and}
  \bibinfo{author}{\bibfnamefont{D.}~\bibnamefont{Bouwmeester}},
  \bibinfo{journal}{Phys. Rev. A} \textbf{\bibinfo{volume}{74}},
  \bibinfo{pages}{013802} (\bibinfo{year}{2006}).

\bibitem{GeorgiadesQuantumSFG1995}
\bibinfo{author}{\bibfnamefont{N.~P.} \bibnamefont{Georgiades}},
  \bibinfo{author}{\bibfnamefont{E.~S.} \bibnamefont{Polzik}},
  \bibinfo{author}{\bibfnamefont{K.}~\bibnamefont{Edamatsu}},
  \bibinfo{author}{\bibfnamefont{H.~J.} \bibnamefont{Kimble}},
  \bibnamefont{and} \bibinfo{author}{\bibfnamefont{A.~S.}
  \bibnamefont{Parkins}}, \bibinfo{journal}{Phys. Rev. Lett.}
  \textbf{\bibinfo{volume}{75}}, \bibinfo{pages}{3426} (\bibinfo{year}{1995}).

\bibitem{HarrisEntangledPhotons2009}
\bibinfo{author}{\bibfnamefont{C.}~\bibnamefont{Belthangady}},
  \bibinfo{author}{\bibfnamefont{S.}~\bibnamefont{Du}},
  \bibinfo{author}{\bibfnamefont{C.-S.} \bibnamefont{Chuu}},
  \bibinfo{author}{\bibfnamefont{G.~Y.} \bibnamefont{Yin}}, \bibnamefont{and}
  \bibinfo{author}{\bibfnamefont{S.~E.} \bibnamefont{Harris}},
  \bibinfo{journal}{Phys. Rev. A} \textbf{\bibinfo{volume}{80}},
  \bibinfo{pages}{031803} (\bibinfo{year}{2009}).

\bibitem{MandelBIphotonInterferometer1991}
\bibinfo{author}{\bibfnamefont{L. J.}~\bibnamefont{Wang}},
  \bibinfo{author}{\bibfnamefont{X. Y.}~\bibnamefont{Zou}}, \bibnamefont{and}
  \bibinfo{author}{\bibfnamefont{L.}~\bibnamefont{Mandel}},
  \bibinfo{journal}{Physical Review A} \textbf{\bibinfo{volume}{44}},
  \bibinfo{pages}{4614} (\bibinfo{year}{1991}).

\bibitem{KorystovBiPhotonInterferometer2001}
\bibinfo{author}{\bibfnamefont{D.~Y.} \bibnamefont{Korystov}},
  \bibinfo{author}{\bibfnamefont{S.}~\bibnamefont{Kulik}}, \bibnamefont{and}
  \bibinfo{author}{\bibfnamefont{A.}~\bibnamefont{Penin}},
  \bibinfo{journal}{Journal of Experimental and Theoretical Physics Letters}
  \textbf{\bibinfo{volume}{73}}, \bibinfo{pages}{214} (\bibinfo{year}{2001}).

\bibitem{MandelBiPhotonInterferometerPRL1991}
\bibinfo{author}{\bibfnamefont{X.~Y.} \bibnamefont{Zou}},
  \bibinfo{author}{\bibfnamefont{L.~J.} \bibnamefont{Wang}}, \bibnamefont{and}
  \bibinfo{author}{\bibfnamefont{L.}~\bibnamefont{Mandel}},
  \bibinfo{journal}{Phys. Rev. Lett.} \textbf{\bibinfo{volume}{67}},
  \bibinfo{pages}{318} (\bibinfo{year}{1991}).

\bibitem{BurlakovBiPhotonInterference1997}
\bibinfo{author}{\bibfnamefont{A.~V.} \bibnamefont{Burlakov}},
  \bibinfo{author}{\bibfnamefont{M.~V.} \bibnamefont{Chekhova}},
  \bibinfo{author}{\bibfnamefont{D.~N.} \bibnamefont{Klyshko}},
  \bibinfo{author}{\bibfnamefont{S.~P.} \bibnamefont{Kulik}},
  \bibinfo{author}{\bibfnamefont{A.~N.} \bibnamefont{Penin}},
  \bibinfo{author}{\bibfnamefont{Y.~H.} \bibnamefont{Shih}}, \bibnamefont{and}
  \bibinfo{author}{\bibfnamefont{D.~V.} \bibnamefont{Strekalov}},
  \bibinfo{journal}{Phys. Rev. A} \textbf{\bibinfo{volume}{56}},
  \bibinfo{pages}{3214} (\bibinfo{year}{1997}).

\bibitem{ZeilingerBiPhotonInterferometer1994}
\bibinfo{author}{\bibfnamefont{T.~J.} \bibnamefont{Herzog}},
  \bibinfo{author}{\bibfnamefont{J.~G.} \bibnamefont{Rarity}},
  \bibinfo{author}{\bibfnamefont{H.}~\bibnamefont{Weinfurter}},
  \bibnamefont{and}
  \bibinfo{author}{\bibfnamefont{A.}~\bibnamefont{Zeilinger}},
  \bibinfo{journal}{Phys. Rev. Lett.} \textbf{\bibinfo{volume}{72}},
  \bibinfo{pages}{629} (\bibinfo{year}{1994}).

\bibitem{RealOrVirtualPhotons1995}
\bibinfo{author}{\bibfnamefont{H.} \bibnamefont{Weinfurter}},
  \bibinfo{author}{\bibfnamefont{T.} \bibnamefont{Herzog}},
  \bibinfo{author}{\bibfnamefont{P. G.} \bibnamefont{Kwiat}},
  \bibinfo{author}{\bibfnamefont{J. G.} \bibnamefont{Rarity}},
  \bibinfo{author}{\bibfnamefont{A.} \bibnamefont{Zeilinger}},
  \bibnamefont{and}
  \bibinfo{author}{\bibfnamefont{M.} \bibnamefont{Zukowski}},
  \bibinfo{journal}{Ann. New York Acad. Sci.} \textbf{\bibinfo{volume}{755}},
  \bibinfo{pages}{61} (\bibinfo{year}{1995}).

\bibitem{EnglertWhichWayInformation1996}
\bibinfo{author}{\bibfnamefont{B.-G.} \bibnamefont{Englert}},
  \bibinfo{journal}{Phys. Rev. Lett.} \textbf{\bibinfo{volume}{77}},
  \bibinfo{pages}{2154} (\bibinfo{year}{1996}).

\endbib

\end{document}